\documentclass{aa}

\usepackage{graphicx}
\usepackage{txfonts}
\usepackage{xcolor}
\usepackage{upgreek}
\usepackage{natbib}
\usepackage{adjustbox}
\usepackage{gensymb}
\usepackage[colorlinks,citecolor=blue,urlcolor=blue,filecolor=blue,linkcolor=blue]{hyperref}
\usepackage{multirow}
\begin{document}
   \title{PSR J0210+5845; An ultra wide binary pulsar with a B6\,V main-sequence star companion}

   \author{E.\ van der Wateren\inst{\ref{astron}, \ref{nijmegen}}
     \and
     C.\ G.\ Bassa\inst{\ref{astron}}  %0000-0002-1429-9010
     \and 
      G.\ H.\ Janssen\inst{\ref{astron}, \ref{nijmegen}}  %0000-0003-3068-3677
      \and
       I.\ V.\ Yanes-Rizo\inst{\ref{iac}, \ref{laguna}}
       \and
       J.\ Casares\inst{\ref{iac}, \ref{laguna}}
       \and\\
       G.\ Nelemans\inst{\ref{nijmegen}, \ref{leuven}, \ref{sron}}
       \and
       B.\ W. Stappers\inst{\ref{jb}}
       \and
       C.\ M.\ Tan\inst{\ref{mcgill1},\ref{mcgill2},\ref{curtin}}
   }

   \institute{ASTRON, Netherlands Institute for Radio Astronomy,
     Oude Hoogeveensedijk 4, 7991 PD Dwingeloo, The
     Netherlands\label{astron}
     \and Department of Astrophysics/IMAPP, Radboud University
     Nijmegen, P.O. Box 9010, 6500 GL Nijmegen, The
     Netherlands\label{nijmegen}
     \and Instituto de Astrof\'isica de Canarias, E-38205 La Laguna, 
     Tenerife, Spain\label{iac}
     \and Departamento de Astrof\'isica, Universidad de La Laguna, 
     E-38206 La Laguna, Tenerife, Spain\label{laguna}
     \and
     Institute of Astronomy, KU Leuven, Celestijnenlaan 200D, 3001 Leuven, Belgium\label{leuven}
    \and
    SRON, Netherlands Institute for Space Research, Niels Bohrweg 4, 2333 CA Leiden, The Netherlands\label{sron}
     \and
     Jodrell Bank Centre for Astrophysics, Department of Physics and 
     Astronomy, University of Manchester, Manchester M13 9PL, UK\label{jb}
     \and
     Department of Physics, McGill University, 3600 rue University, Montr\'eal, QC H3A 2T8, Canada\label{mcgill1}
     \and
     The Trottier Space Institute at McGill, 3550 rue University, Montr\'eal, QC H3A 2A7, Canada\label{mcgill2}
     \and 
     International Centre for Radio Astronomy Research, Curtin University, Bentley, WA 6102, Australia\label{curtin}
   }

   \date{Received \today; accepted \today}

   \abstract{We report on radio timing observations of PSR\,J0210+5845 which reveal large deviations from typical pulsar spin-down behaviour. We interpret these deviations as being due to binary motion around the $V=13.5$ star 2MASS\,J02105640$+$5845176, which is coincident in celestial position and distance with the pulsar. Archival observations and new optical spectroscopy identify this star as a B6\,V star with a temperature of $T_\mathrm{eff}\approx 14\,000$\,K and a mass of $M_\mathrm{c}= 3.5$ to $3.8$\,M$_\odot$, making it the lowest mass main-sequence star known orbiting a non-recycled pulsar. We found that the timing observations constrain the binary orbit to be wide and moderately eccentric, with an orbital period of $P_\mathrm{b}=47^{+40}_{-14}$\,yr and eccentricity $e=0.46^{+0.10}_{-0.07}$. We predict that the next periastron passage will occur between 2030 and 2034. Due to the low companion mass, we find that the probability for a system with the properties of PSR\,J0210+5845 and its binary companion to survive the supernova is low. We show that a low velocity and fortuitously directed natal kick is required for the binary to remain bound during the supernova explosion, and argue that an electron-capture supernova is a plausible formation scenario for the pulsar.}


\maketitle

\section{Introduction}
Within the population of about 3300 radio pulsars presently known \citep{Manchester05}, there exists a distinct sub-population of six binary systems where a normal (non-recycled) pulsar, orbits a massive stellar companion. These binary systems have eccentric orbits with orbital periods on the order of months or years \citep{Kaspi96,  Lorimer06, Shannon14} or even decades \citep{Lyne15}. The secondary stars are O or B stars with masses exceeding 8\,M$_\odot$ \citep{Bell95, Lyne15}, and for those systems located in our Galaxy, they have low Galactic latitudes. Table \ref{tab:hmbps} summarises the main properties of the six pulsar/massive star binaries known to date.

Most of the systems exhibit significant interaction between the pulsar and the stellar wind or disc of the massive stellar companion \citep{Andersen23}, leading to variations in scattering and dispersion \citep{Madsen12}, eclipses of the pulsar emission \citep{Wang04}, and/or X-ray or gamma-ray emission \citep{Aharonian05}. As a result, these systems serve as exceptional laboratories for investigating binary interactions in stellar disks and winds. They are thought to be possible progenitors of double neutron star systems \citep{Johnston94}.

The progenitors of these pulsar/massive star binaries are systems composed of two massive main-sequence stars. In such systems, no  stellar interaction is necessarily required, although it is possible that mass transfer has occurred in systems with small orbital periods. When the primary star undergoes supernova to form the neutron star, it sheds a large portion of its envelope \citep{Matzner99}. If a binary system loses more than half of its mass, the system is usually disrupted, because the orbital energy surpasses the binding energy \citep{Hills83}. Furthermore, asymmetries in the supernova explosion can impart a natal kick to the newborn neutron star, and depending on the velocity (speed and direction) of the kick with respect to the orbital velocity, the binary binding energy can be further altered, possibly counteracting the disruption of the binary or vice versa. In those cases where the binary remains bound, the neutron star will orbit the unaltered massive main-sequence star in a wide orbit with high eccentricity \citep[e.g.][]{Brandt95}.

In 2017, PSR\,J0210+5845 was discovered as part of the LOFAR Tied-Array All-Sky Survey (LOTAAS), the LOFAR pulsar survey of the northern hemisphere \citep{Sanidas19}. \citet{Tan20} showed that initial timing observations revealed significant timing residuals that they argue could be caused by timing noise intrinsic to the pulsar. In this study, we show that the timing residuals are the result of binary motion with an $M_\mathrm{c}= 3.5$ to $3.8$\,M$_\odot$ companion in a moderately eccentric, long period orbit. The PSR\,J0210+5845 system has the lowest-mass binary companion among known pulsar/massive star binaries, posing a challenge for its survival of the supernova explosion that formed the neutron star.  

We present continued timing observations in Section\,\ref{sec:timing} that show that the spindown of PSR\,J0210+5845 can be modelled by several higher-order spin-frequency derivatives. We find that a $V\sim13.5$ star is coincident with the pulsar timing position, discuss its properties as the optical counterpart to PSR\,J0210+5854, and identify it as the binary companion of the pulsar in Section\,\ref{sec:counterpart}. In Section\,\ref{sec:orbital_model}, we use the spin frequency derivatives to obtain orbital constraints and investigate formation scenarios in Section\,\ref{sec:formation}. We discuss and conclude in Section\,\ref{sec:discussion}.

\begin{table*}[htbp]
    \centering
    \begin{tabular}{lrrrrrrr}
    \hline
        PSR name & $P$ & $P_\mathrm{b}$ & $e$ & $M_\mathrm{c}$ & $l$ & $b$ & Ref.\\
         & (s) & (d) & & (M$_\odot$) & & & \\
        \hline
        J0045$-$7319 & 0.926 & 51 & 0.81 & 8.8 & $303\fdg51$ & $-43\fdg80^\star$ & [1, 2] \\
        B1259$-$63 & 0.048 & 1237 & 0.87 & 20 & $304\fdg18$ & $-0\fdg99$ & [3, 4]\\
        J1638$-$4725 & 0.764 & 1941 & 0.96 & $8^\dagger$ & $337\fdg36$ & $-0\fdg30$ & [5] \\
        J1740$-$3052 & 0.570 & 231 & 0.58 & 20 & $357\fdg81$ & $-0\fdg13$ & [6, 7]\\
        J2032+4127 & 0.143 & 16835 & 0.96 & 15 & $80\fdg22$ & $+1\fdg03$ & [8]\\
        J2108+4516 & 0.577 & 269 & 0.09 & 17.5 -- 23 & $87\fdg34$ & $-1\fdg63$ & [9]\\
        \hline
    \end{tabular}
    \caption{The six currently known pulsar/massive star binaries and their spin period ($P$), orbital period ($P_\mathrm{b}$), eccentricity ($e$), companion mass ($M_\mathrm{c}$), Galactic longitude ($l$) and latitude ($b$). The references used are: [1] \citet{Kaspi96}, [2] \citet{Bell95}, [3] \citet{Shannon14}, [4] \citet{Johnston94}, [5] \citet{Lorimer06}, [6] \citet{Bassa+11}, [7] \citet{Madsen12}, [8] \citet{Lyne15}, and [9] \citet{Andersen23}. $^\star$ PSR\,J0045$-$7319 is located in the Small Magellanic Cloud, hence the large Galactic latitude. $^\dagger$ The companion of J1638$-$4725 has not been identified. The mass estimate is the median mass calculated from the orbital period and projected-semi major axis, assuming an inclination of $60^\circ$.}\label{tab:hmbps}
\end{table*}

\section{Radio timing}\label{sec:timing}
The timing ephemeris of PSR\,J0210+5845 obtained by \citet{Tan20}, modelling the position, the spin period and its derivative, and the dispersion measure (DM), uses LOFAR observations obtained between December 2017 and December 2018. We extended that timing ephemeris with observations from 2019 to June 2022. All observations of PSR\,J0210+5845 were obtained with the same observational setup. We  used the high-band antennas (HBAs) of the LOFAR core stations, recording dual-polarisation Nyquist sampled complex voltages for 400 subbands of 0.195\,MHz bandwidth between 110 to 188\,MHz. We followed the analysis procedure outlined in \citet{vanderWateren23}, where we used the LOFAR pulsar pipeline \citep{Kondratiev16} to coherently dedisperse and fold the observations with \textsc{dspsr} \citep{VanStraten11} to create pulse profiles in the \textsc{psrfits}\footnote{\url{https://psrchive.sourceforge.net}} format with 0.195\,MHz channels and 5-s subintegrations. The majority of radio frequency interference (RFI) was automatically removed using \textsc{clfd}\footnote{\url{https://github.com/v-morello/clfd}} \citep{Morello19}, followed by a manual inspection and RFI removal with the \textsc{psrzap} tool from the \textsc{psrchive} software suite \citep{Hotan04} where needed.

We used the timing model from \citet{Tan20} to fold and dedisperse all observations, which were then fully averaged in time. To obtain a better constraint on the DM, the observations were split into two subbands with centre frequencies 129 and 167\,MHz and both subbands were fully averaged in frequency. We combined the observations to one high signal-to-noise profile of the full bandwidth, to which we modelled an analytical template profile as the sum of three von Mises functions using \textsc{paas}. The averaged observations were cross-correlated with the template to obtain times-of-arrival (TOAs) with \textsc{pat}. 

\begin{figure}[!h]
  \centering
  \includegraphics[width=\columnwidth]{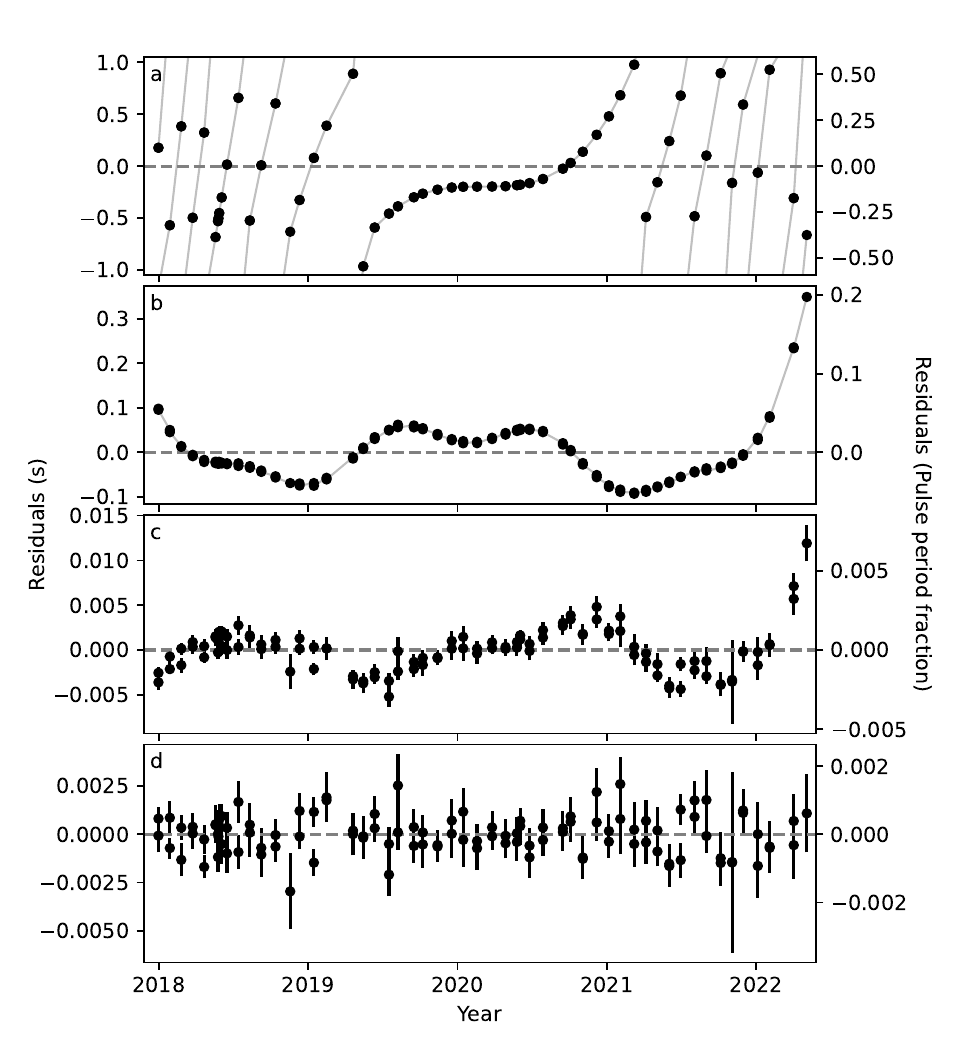}
  \caption{The residuals from timing with one (a), two (b) three (c), and four (d) spin frequency derivatives. The residuals from fitting the model in Table\,\ref{tab:timing} fitting up to and including five spin frequency derivatives are very similar to the residuals shown here.}
    \label{fig:residuals}
\end{figure}

With \textsc{pintk}, the interactive module of the pulsar timing package \textsc{PINT}\footnote{\url{https://nanograv-pint.readthedocs.io/en/latest}} \citep{Luo21} (v0.9.3), a new timing ephemeris was constructed modelling the celestial position ($\alpha_\mathrm{J2000}$, $\delta_\mathrm{J2000}$), spin frequency $f$, spin frequency derivative $\dot{f}$, and the DM. As indicative in Fig.\,\ref{fig:residuals}, higher-order spin frequency derivatives were required to properly model the spindown behaviour of PSR\,J0210+5854. We sequentially added spin frequency derivatives beyond $\dot{f}$ to improve the timing ephemeris and obtained significant measurements for the second, third, and fourth spin frequency derivatives $\ddot{f}$, $\dot{\ddot{f}}$, and $\ddot{\ddot{f}}$. Fitting a timing ephemeris that also included a fifth spin frequency derivative in the fit resulted in a value that was not significant (less than $3\sigma$ significance). To assess the timing noise using the method from \citet{Arzoumanian94}, we fitted $f$, $\dot{f}$, and $\ddot{f}$, keeping higher-order frequency derivatives at zero, over a $10^8$-s segment of the data to calculate $\Delta_8 = \log_{10} \Big((6f)^{-1}|\ddot{f}|t^3\Big)$, with $t=10^8\,$s. The obtained value of $\Delta_8=1.54$ significantly exceeds the expected value for intrinsic timing noise of $\Delta_8=-1.3$ based on the relation from \citet{Hobbs10}, which suggests that the higher-order frequency derivatives are not primarily produced by intrinsic timing noise. 

The higher-order spin frequency derivatives up to and including the fifth spin frequency derivative were included in the timing ephemeris. Even though the fifth spin frequency derivative was not significant, it was included due to the informative nature of its uncertainty, particularly in terms of the scale for this parameter, which was used in Section \ref{sec:orbital_model}. The timing model was subsequently refitted for all parameters, resulting in the timing ephemeris shown in Table\,\ref{tab:timing} and the timing residuals in Fig.\,\ref{fig:residuals}.

To check that there are no unexpected covariances between the parameters in the timing ephemeris, we performed a Bayesian analysis using the Markov Chain Monte Carlo (MCMC) fitter from \texttt{PINT}. The standard \texttt{PINT} approach was used by taking normally distributed priors based on the standard \texttt{PINT} fitting results and refitting all parameters in the timing ephemeris using the MCMC fitter. The parameters and their uncertainties are consistent between fitting methods, and covariances are generally low except for pairs of odd and even spin frequency derivatives, as expected for a Taylor series.

To investigate for variations in dispersion, we divided the TOAs into segments, each spanning 200 days, which we independently refitted for DM. This analysis revealed a maximum variation of 0.38(49)\,pc\,cm$^{-3}$, with the uncertainty denoting the average error associated with the separately fitted DMs. Refitting the full data set for a time derivative of the DM resulted in $\dot{\mathrm{DM}}=-0.0006(15)$\,pc\,cm$^{-3}$\,yr$^{-1}$. Hence, we conclude that PSR\,J0210+5845 exhibits no significant DM variations in our dataset.

Coincident with the timing position of PSR\,J0210+5845 is an optical star for which the position and proper motions are documented in the {\it Gaia} DR3 catalogue \citep{Gaia23} and which, in Section\,\ref{sec:counterpart}, we identify as the binary companion of the pulsar. We refitted the spin frequency, higher-order derivatives of the spin frequency and the DM,  incorporating the position and proper motions from the optical counterpart of PSR\,J0210+5845 as documented in the {\it Gaia} DR3 catalogue \citep{Gaia23}. The resulting timing model remained consistent with the model in which the position was fitted and no proper motion was assumed. See Section\,\ref{sec:counterpart} for more information on the optical counterpart.

\begin{table*}
  \centering
  \caption{The timing parameters for PSR\,J0210+5845. Two timing ephemerides are fitted, one where the celestial position is fitted together with five spin frequency derivatives and the dispersion measure, and one where the celestial position and proper motions are fixed to those of {\it Gaia} DR3. The {\it Gaia} positions are consistent with the positions from radio timing at the same epoch of position measurement. The higher-order frequency derivatives and DMs are consistent at the epoch of frequency determination. The figures in parentheses are the nominal $1\sigma$ uncertainties in the least significant digits quoted, which have been multiplied by the square root of the reduced $\chi^2$.}
  \label{tab:timing}
  \begin{tabular}{lrr}
    \hline
    
    \hline
    \multicolumn{3}{c}{Data set and assumptions} \\
%    Pulsar name & \multicolumn{2}{c}{J0210+5845} \\
    MJD range & \multicolumn{2}{c}{58118--59704} \\
    Data span (yr) & \multicolumn{2}{c}{4.34} \\
    Number of TOAs & \multicolumn{2}{c}{110} \\
    Clock correction procedure & \multicolumn{2}{c}{TT(TAI)} \\
    Solar system ephemeris model & \multicolumn{2}{c}{DE436}\\
    Units & \multicolumn{2}{c}{TDB}\\
   \hline
    \multicolumn{3}{c}{Measured quantities} \\
    & Fitting for astrometry   & {\it Gaia} DR3 astrometry \\
    Epoch of position measurement (MJD) &  58910.0 & 57388.5\\
    Right ascension, $\alpha_\mathrm{J2000}$  & $02^\mathrm{h}10^\mathrm{m}56\fs416(9)$ & $02^\mathrm{h}10^\mathrm{m}56\fs409999$\\ 
    Declination, $\delta_\mathrm{J2000}$ &  $+58\degr45\arcmin17\farcs74(8)$ & $+58\degr45\arcmin17\farcs718237$\\
    Proper motion in RA, $\mu_\alpha \cos \delta$ (mas yr$^{-1}$) & $\ldots$ & $-$1.116\\
    Proper motion in Dec., $\mu_\delta$ (mas yr$^{-1}$) & $\ldots$ & $-$0.495\\
    Epoch of frequency determination (MJD)& 58910.0 & 58910.0 \\
    Spin frequency, $f$ (s$^{-1}$)& 0.566181225686(6) &  0.566181225686(6) \\
    First derivative of spin frequency, $\dot{f}$ (s$^{-2}$) & $-$4.86224(6)$\times10^{-14}$ & $-$4.86223(6)$\times10^{-14}$\\
    Second derivative of spin frequency, $\ddot{f}$ (s$^{-3}$)& $-$1.1628(3)$\times 10^{-22}$ & $-$1.1628(3)$\times 10^{-22}$  \\
    Third derivative of spin frequency, $\dot{\ddot{f}}$ (s$^{-4}$) & $-$4.96(4)$\times 10^{-31}$  & $-$4.96(4)$\times 10^{-31}$ \\
    Fourth derivative of spin frequency, $\ddot{\ddot{f}}$ (s$^{-5}$) & $-$2.47(13)$\times 10^{-39}$  & $-$2.47(12)$\times 10^{-39}$ \\
    Fifth derivative of spin frequency, $\dot{\ddot{\ddot{f}}}$ (s$^{-6}$) & $-$1.9(2.0)$\times 10^{-47}$ & $-$1.5(1.9)$\times 10^{-47}$\\
    Dispersion measure, DM (pc\,cm$^{-3}$) & 76.7895(18) & 76.7894(18) \\
    Weighted rms timing residual ($\upmu$s) & 860.7 & 864.5 \\
    Reduced $\chi^2$ value & 0.89 & 0.90\\
    \hline
  \end{tabular}
\end{table*}

\begin{table}[htbp]
    \centering
    \tiny
    \renewcommand{\arraystretch}{1.5} % Increase row spacing
    \begin{tabular}{rrrl}
    \hline
        $d$ & $M_\mathrm{c}$ & $T_\mathrm{eff}$ & Ref.\\
         (kpc) & (M$_\odot$) & (K) &   \\
        \hline
        $2.5(5)^*$ & - & - & \citet{Cordes02} \\
$2.0(4)^*$ & - & - & \citet{Yao17} \\ 
$2.31^{+0.27}_{-0.20}$ & $2.78^{+0.48}_{-0.83}$ & $11\,457^{+972}_{-4298}$ & \citet{Anders19} \\
- & - & 7840(253) & \cite{Xiang19} \\
$1.95^{+0.04}_{-0.07}$ & $1.69^{+0.10}_{-0.04}$ & $8261^{+285}_{-94}$ & \citet{Queiroz20}\\
$2.60^{+0.11}_{-0.10}$ & - & - & \citet{Bailer-Jones21} \\
$2.51^{+0.05}_{-0.14}$ & $3.22^{+0.40}_{-0.62}$ & $12\,260^{+1408}_{-1311}$ & \citet{Anders22} \\ 
$2.2070^{+0.0042}_{-0.0011}$ & $3.10^{+0.07}_{-0.05}$ & $10\,930^{+5}_{-28}$ & \citet{Gaia23} \\ 
 - & - & 14\,279(603) & \citet{Xiang22} \\ 
 - & $3.5$ to $3.8$ & 14275(133) & This paper \\
        \hline
    \end{tabular}
    \caption{The distance $d$, mass $M_\mathrm{c}$, and effective temperature $T_\mathrm{eff}$ of 2MASS\,J02105640+584517 as documented in different catalogues. The uncertainties are 68\% confidence intervals as upper and lower bounds or $1\sigma$ uncertainties in parentheses. Distances denoted with $^*$ are derived from the location and DM of the pulsar using the NE2001 \citep{Cordes02} and YMW16 \citep{Yao17} Galactic electron density models. }\label{tab:companion}
\end{table}

\section{Optical counterpart}\label{sec:counterpart}
A star ($V=13.5$, $B - V = 0.47$; \citealt{Henden15}) is located near the radio timing position of PSR\,J0210+5845. This star is present in many catalogues -- for the remainder of the paper, we will refer to it as 2MASS\,J02105640+5845176 \citep{Skrutskie06}. The {\it Gaia} DR3 \citep{Gaia23} astrometric solution provides a proper motion of $\mu_\alpha \cos{\delta}=-1.116(12)$\,mas\,yr$^{-1}$ and $\mu_\delta=-0.495(15)$\,mas\,yr$^{-1}$. At the epoch of the pulsar timing ephemeris, the position of this star is $\alpha_\mathrm{ICRS}=02^\mathrm{h}10^\mathrm{m}56^\mathrm{s}.4094$, $\Delta \delta_\mathrm{ICRS} = +58\degree45\arcmin17\farcs716$. This position is offset from the pulsar timing position by $\Delta \alpha=0\farcs06(7)$ in right ascension and $\Delta\delta= 0\farcs02(8)$ in declination and hence coincident with the pulsar position as measured through timing.

The parallax of this star has been measured at $\varpi = 0.354(15)$\,mas in {\it Gaia} DR3. The {\it Gaia} GSP-Phot and FLAME modelling combine this parallax with GAIA photometry and stellar models to estimate a distance of $d=2.21$\,kpc, a nominal effective temperature of $T_\mathrm{eff}=10930$\,K, and a mass of $M_\mathrm{c}=3.10$\,M$_\odot$ \citep{Gaia23}. Similar stellar properties were obtained with the \texttt{StarHorse} Bayesian isochrone-fitting software \citep{Queiroz18} by combining the GAIA astrometric and photometric measurements with photometry from Pan-STARRS1, 2MASS, and AllWISE. Results based on {\it Gaia} DR2 are presented in \citet{Anders19} and updated using {\it Gaia} EDR3 in \citet{Anders22}. These results are shown in Table\,\ref{tab:companion}.

Low-resolution spectroscopy of 2MASS\,J02105640+5845176 has been obtained as part of the LAMOST survey, which initially classified the star as an A1\,V star and estimates a radial velocity of $-7.6(5)$\,km\,s$^{-1}$ \citep{Xiang22}. Stellar parameters derived from this spectrum vary between analysis methods, with $T_\mathrm{eff}=7840(253)$\,K obtained by \citet{Xiang19} from LAMOST DR5 and $T_\mathrm{eff}=14279(603)$\,K by \citet{Xiang22} from LAMOST DR6. \citet{Queiroz20} used the LAMOST DR5 stellar parameters for the lower effective temperature solution from \citet{Xiang19} with the \texttt{StarHorse} software to obtain lower mass and distance estimates of 2MASS\,J02105640+5845176 ($d=1.95$\,kpc, $M_\mathrm{c}=1.69$\,M$_\odot$) compared to the photometric results (see Table \,\ref{tab:companion}). The higher effective temperature obtained in \citet{Xiang22} is consistent with the mass and temperature estimates derived from astrometry and photometry.

To resolve this discrepancy in the effective temperature of 2MASS\,J02105640$+$5845176, we obtained four spectroscopic observations of 2MASS\,J02105640$+$5845176 during morning twilight on August 16 and 17, 2022 with the Intermediate Dispersion Spectrograph (IDS) at the 2.5\,m Isaac Newton Telescope on La Palma. The R900V grating was used with 600-s exposures on the RED+2 detector. The seeing varied between $1\farcs3$ and $2\farcs1$. We used the $0\farcs974$ slit, covering the wavelength range of 3800 to 5400\,\AA\ at 0.70\,\AA\,pix$^{-1}$, providing a resolution of $R\sim3000$. The observations were bias subtracted and spectra were extracted using the method described by \citet{Hynes02}. Arc-lamp exposures taken prior to each observation of 2MASS\,J02105640$+$5845176 were used for the wavelength calibration. Radial velocities and spectral properties were determined by fitting the observed spectra against normalised model spectra from \citet{Munari05}. The model spectra were convolved with a truncated Gaussian to decrease their resolution of $R=20\,000$ to that of the observations.

We found that the barycentred radial velocities are consistent with a mean velocity of $v=-9(3)$\,km\,s$^{-1}$ ($3.2$\,km\,s$^{-1}$ rms around the mean). The observed spectrum is best represented by the models with Solar metallicity ($[\mathrm{M}/\mathrm{H}]=0$) with an effective temperature $T_\mathrm{eff}=14\,275(133)$\,K, surface gravity $\log (g/cgs) >4.13$ ($2\sigma$) and rotationally broadened by $v_\mathrm{rot}\sin i=73(12)$\,km\,s$^{-1}$. At this effective temperature and surface gravity, 2MASS\,J02105640$+$5845176 can be classified as a B6\,V star \citep{Pecaut13}.

The PARSEC stellar evolution models by \citet{Bressan12,Tang14} and \citet{Chen14,Chen15} predict a mass of $M_\mathrm{c}=3.59(5)$\,M$_\odot$ and absolute V-band magnitude of $M_V=0.28(4)$ for an $[\mathrm{M}/\mathrm{H}]=0$ main-sequence star with $T_\mathrm{eff}=14\,275(133)$\,K at an age of 10\,Myr. At 40\,Myr, the mass and absolute V-band magnitude are $M_\mathrm{c}=3.72(6)$\,M$_\odot$ and $M_V=0.06(4)$. The \citet{Green19} Galactic extinction map predicts $E_{g-r}=0.38(2)$ to $0.40(2)$\,mag for distances from 1.9 to 3.0\,kpc. For these reddening values, the $R_V=3.1$ extinction coefficients from \citet{Schlafly11} yield $A_V=1.15$ to 1.21\,mag. Combined with the observed magnitude of $V=13.507$ \citep{Henden15} this yields distances of $d=2.54$\,kpc to 2.80\,kpc. From these observations, we conclude that the A1\,V stellar classification and the $T_\mathrm{eff}=7840$\,K estimate by \citet{Xiang19} is in error and that 2MASS\,J02105640$+$5845176 is a $M_\mathrm{c}= 3.5$ to $3.8$\,M$_\odot$, $T_\mathrm{eff}\sim14\,000$\,K B6\,V star at a distance of $d=2.5$ to 2.8\,kpc.

At the location and $\mathrm{DM}$ of PSR J0210+5845, the NE2001 \citep{Cordes02} and YMW16 \citep{Yao17} Galactic electron density models predict distances of 2.52 and 1.95\,kpc, respectively. Given the typically 20\% uncertainties on DM-derived distances, these distances are consistent with the distances derived from the {\it Gaia} parallax measurement and the photometric and spectroscopic constraints.

The uncertainty in the tie between the pulsar position and the {\it Gaia} astrometry is dominated by the uncertainty in the former. The 99\% confidence error ellipse on the pulsar timing position has an area of 0.053 sq.\ arcsec, while the {\it Gaia} DR3 object density towards PSR\,J0210+5854 is only 18.4 stars per sq.\ arcmin. Hence, the probability of finding an unrelated star from the {\it Gaia} DR3 catalogue in the error ellipse of PSR\,J0210+5854 is low $p=2.6\times10^{-4}$. Given this low chance coincidence, and the consistent distance estimates, we will consider the optical counterpart 2MASS\,J02105640+5845176 as the binary companion to PSR\,J0210+5854 for the remainder of the paper.

Finally, we note that based on light curves from ZTF \citet{Bellm19}, \citet{Chen20} identified 2MASS\,J02105640+5845176 as a periodic variable, with 0.042\,mag variations in $r$-band on a 1.1248\,d period, and classified it as a variable of the RS\,CVn type. This classification is inconsistent with the spectral type obtained from spectroscopy and the mass and temperature estimates from astrometry and photometry. Additionally, our optical spectroscopy rules out radial velocity variations larger than 3\,km\,s$^{-1}$ over a 24\,h period, ruling out the RS\,CVn classification. We consider it more likely that 2MASS\,J02105640+5845176 is a slowly pulsating B-star, a variable type that produces photometric variability of similar periodicity and amplitude in stars of spectral types ranging from B2 to B9 \citep{Waelkens91, Fedurco20}.

\begin{figure*}
  \centering
  \includegraphics[width=\textwidth]{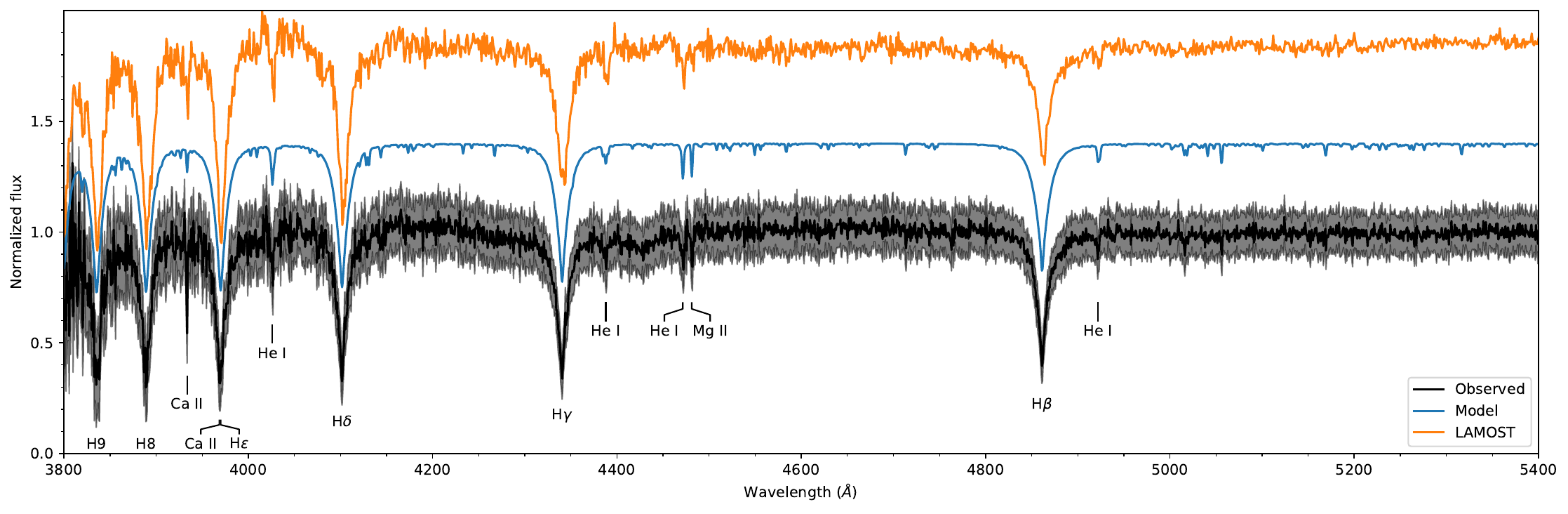}
  \caption{The normalised optical spectrum of 2MASS\,J02105640+5845176 (black) as observed with the Intermediate Dispersion Spectrograph at the Isaac Newton Telescope. The best fitting model spectrum from \citet{Munari05} is plotted in blue (shifted by +0.4 units vertically) and has $T_\mathrm{eff}=14\,000$\,K, $\log g=4.5$ cgs, $[\mathrm{M}/\mathrm{H}=0$ and $v_\mathrm{rot}\sin i=75$\,km\,s$^{-1}$. The model spectrum has been convolved with the response of the slit to match the resolution of the observed spectrum. The LAMOST DR7 spectrum from \citet{Xiang22} is shown in orange (shifted by +0.8 units vertically). Prominent absorption lines are indicated.}
    \label{fig:spectrum}
\end{figure*}

\section{Orbital constraints}\label{sec:orbital_model}
To investigate if binary motion between PSR\,J0210+5845 and 2MASS\,J02105640+5845176 can explain the observed higher-order spin frequency derivatives in the timing of PSR\,J0210+5845, we used the method from \citet{Bassa16}. This method is based on the derivations of \citet{Joshi97} and uses a Keplerian orbit to compute time derivatives of the line-of-sight position of the pulsar to predict spin frequency derivatives. We consider it unlikely that PSR\,J0210+5845 and 2MASS\,J02105640+5845176 are in an unbound, hyperbolic, orbit, as the time scale for gravitational interaction will be extremely short compared to the lifetime of the pulsar. Hence, we modelled the observed spin frequency derivatives with a bound Keplerian orbit described by the orbital period $P_\mathrm{b}$, projected semi-major axis $x$, eccentricity $e$, argument of perigee $\omega$, and true anomaly $\nu$. 

For comparison of the observed spin frequency derivatives with predicted values from the Keplerian model, we used the first to fifth spin frequency derivatives. We implicitly assumed that the second and higher-order spin frequency derivatives are entirely dominated by orbital motion, but need to correct the observed first-order spin frequency derivative $\dot{f}_\mathrm{obs}$ for the unknown intrinsic spin frequency derivative $\dot{f}_\mathrm{int}$ due to spin down. The contributions to $\dot{f}_\mathrm{obs}$ from the Shklovskii \citep{Shklovskii70} effect and the differential and Galactic acceleration \citep[e.g.][]{Nice95} are at least six orders of magnitude smaller than $\dot{f}_\mathrm{obs}$ and are therefore neglected. From the ATNF Pulsar Catalogue \footnote{\url{http://www.atnf.csiro.au/research/pulsar/psrcat}} (version 1.67, \citealt{Manchester05}), we found that the intrinsic spin frequency derivative distribution of normal, non-recycled pulsars (those with spin period $P>0.02$\,s) can be described by a log-normal distribution of the form $\log_{10} {-\dot{f}_\mathrm{int}}=-14.3(1.3)$ (for $\dot{f}_\mathrm{int}$ in units of s$^{-2}$).

We performed a Monte Carlo simulation to predict spin frequency derivatives for 100\,000 randomly sampled Keplerian orbits. Samples of $e$, $\omega$, and $\nu$ were drawn from uniform distributions between $0 \leq e < 1$, $0\degree \leq \omega < 360\degree$, and $0\degree \leq \nu < 360\degree$, respectively. Using the drawn samples, we used the equations from \citet{Bassa16} and solved for $P_\mathrm{b}$ and $x$. Random values for the intrinsic spin frequency derivative $\dot{f}_\mathrm{int}$ were drawn from the log-normal distribution to correct the observed spin frequency derivative and obtain the contribution due to orbital motion on $\dot{f}$. For each set of parameters, we calculated the spin frequency derivatives, which we compared to the observed values from the timing analysis in Section\,\ref{sec:timing}. We only retained parameter sets for which the predicted spin frequency derivatives were within $3\sigma$ of the observed values and for which the orbital inclination $i$ was consistent with $\sin i<1$ for a 1.4\,M$_\odot$ pulsar and a $M_\mathrm{c}=3.6$\,M$_\odot$ binary companion. 

\begin{figure*}
  \centering
  \includegraphics[width=\textwidth]{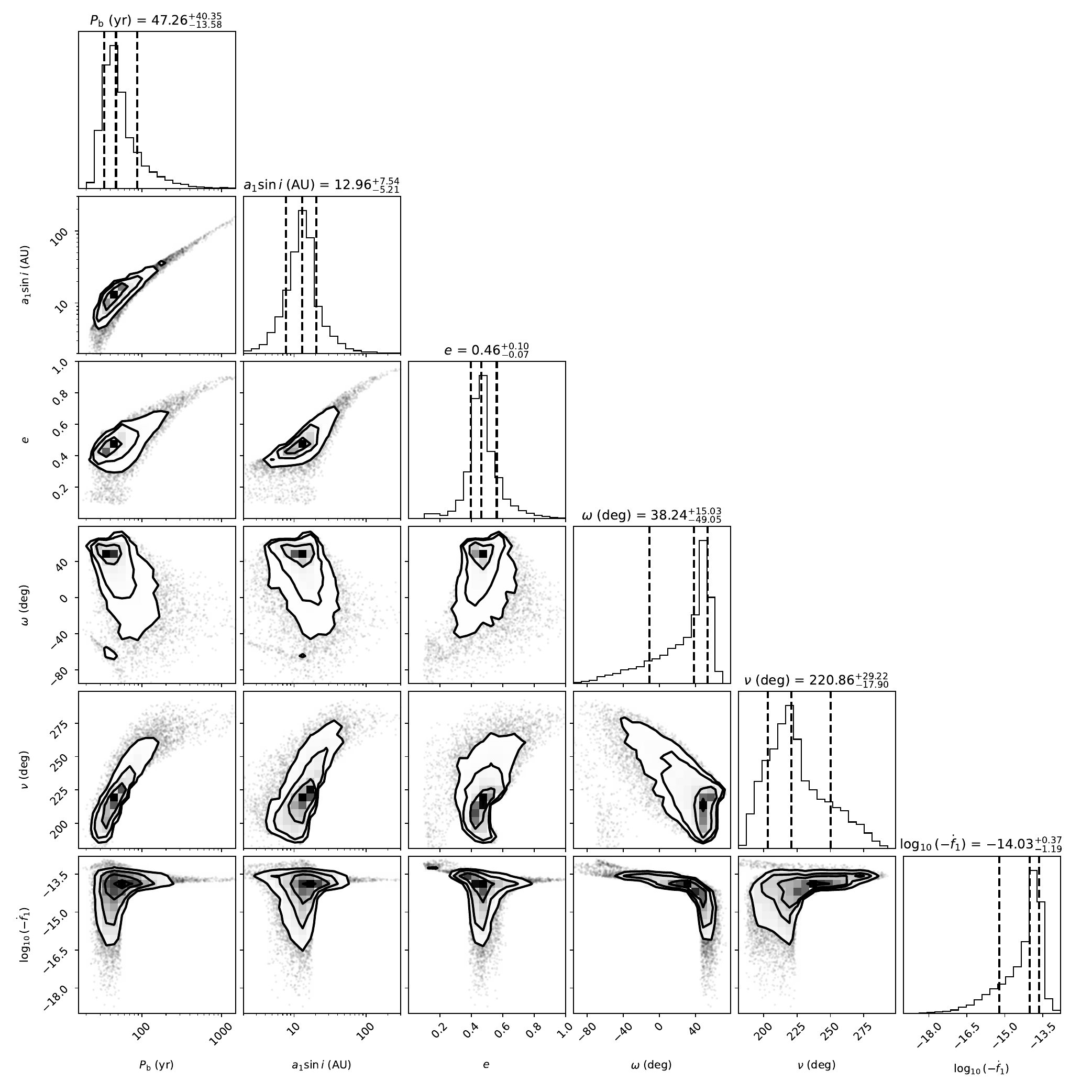}
  \caption{The parameter distributions of  the orbital period ($P_\mathrm{b}$), the projected semi-major axis ($x$), the eccentricity ($e$), the argument of perigee ($\omega$), the true anomaly ($\nu$), and the intrinsic spin frequency derivative ($\dot{f}_\mathrm{int}$) resulting from the Monte Carlo simulation assuming $M_\mathrm{c}=3.6\,$M$_\odot$. The median values and the 68\% confidence intervals for each parameter are displayed on top of the columns.}
    \label{fig:corner_grid}
\end{figure*}

The results of the Monte Carlo simulation are presented in Fig.\,\ref{fig:corner_grid} and they indicate that the observed spin frequency derivatives can be explained by a wide and moderately eccentric orbit with $P_\mathrm{b}=47^{+40}_{-14}$ yr, $x=12.9^{+7.5}_{-5.2}$\,AU and $e=0.46^{+0.10}_{-0.07}$ (68\% confidence intervals). The orbital inclination is constrained to $i=52(18)\degr$ and with a true anomaly of $\nu\sim221\degr$, the binary system had its previous apastron passage between 1988 and 2016 ($T_\mathrm{ap}=2009^{+7}_{-20}$) while the next periastron passage is predicted to occur between 2030 and 2034 ($T_\mathrm{per}=2032.1^{+1.7}_{-1.3}$). At periastron, the distance between the binary components will be $q=11.7^{+5.0}_{-1.5}$\,AU. By correcting for the orbital motion of PSR\,J0210+5845, we obtained an intrinsic spin-down $\dot{f}_\mathrm{int} = -0.6^{+1.7}_{-0.5}\times10^{-14}\,$s$^{-2}$, constraining the characteristic age to $\tau_\mathrm{c}=1.6^{+13}_{-1.1}$\,Myr.

If we assume a lower companion mass of 1.7\,M$_\odot$, we obtain a very similar estimate for $\nu$ and slightly lower estimates for $P_\mathrm{b}$, $x$, $e$, $\omega$, and $\dot{f}_\mathrm{int}$, but the values are consistent with those from $M_\mathrm{c}=3.6\,$M$_\odot$.

\section{Formation scenarios}\label{sec:formation}
The formation of high-mass binary pulsars follows that of high-mass X-ray binaries and the first-born neutron star in double neutron star systems, where the more massive primary of a binary consisting of O and/or B stars undergoes a supernova explosion to form the neutron star \citep[e.g.][and references therein]{Brandt95,Tauris17}. Alternatively, an episode of mass transfer from the primary to the secondary can lead the primary to expel most of its envelope, leaving a Helium star that can explode as a Type Ib/c supernova \citep{Eldridge08}. 

PSR\,J0210+5845 poses a challenge for these formation scenarios as, compared to other high-mass binary pulsars, it has a relatively low companion mass of $\sim3.6$\,M$_\odot$. Since, in the absence of mass transfer, the progenitor of the pulsar in PSR\,J0210+5845 must have had a mass larger than $\ga8$\,M$_\odot$ to form a $\sim1.4$\,M$_\odot$ neutron star \citep{Woosley02}, at least 56\% of the mass in the binary system would have been lost in a direct supernova explosion. As this exceeds the 50\% limit above which the binary is disrupted \citep{Hills83}, a natal kick imparted on the neutron star during the supernova explosion would be required to keep the binary system bound. 

To investigate the kick velocities required for the system to survive the mass loss in the supernova explosion, we used the formalism by \citet{Brandt95} to calculate post-supernova orbits. As input for the direct supernova channel, we assumed as initial conditions a $M_\mathrm{p}=8$ to 12\,M$_\odot$ neutron star progenitor and a $M_\mathrm{c}=3.6$\,M$_\odot$ secondary in a wide orbit (initial orbital periods larger than $P_\mathrm{b,ini}>1500$\,d) to ensure it does not fill its Roche lobe before the star explodes \citep{Klencki22}. We assumed that the initial orbit is circularised due to tides from the neutron star progenitor. For these parameters, the \citet{Brandt95} equations predict a maximum kick velocity of $v_\mathrm{kick}<96$\,km\,s$^{-1}$, above which the system will not remain bound. Hence, random kick velocities were chosen with magnitudes up to this limit and we used the standard assumption that the distribution in kick directions is isotropic.

\begin{figure*}
    \centering
    \includegraphics[width=\textwidth]{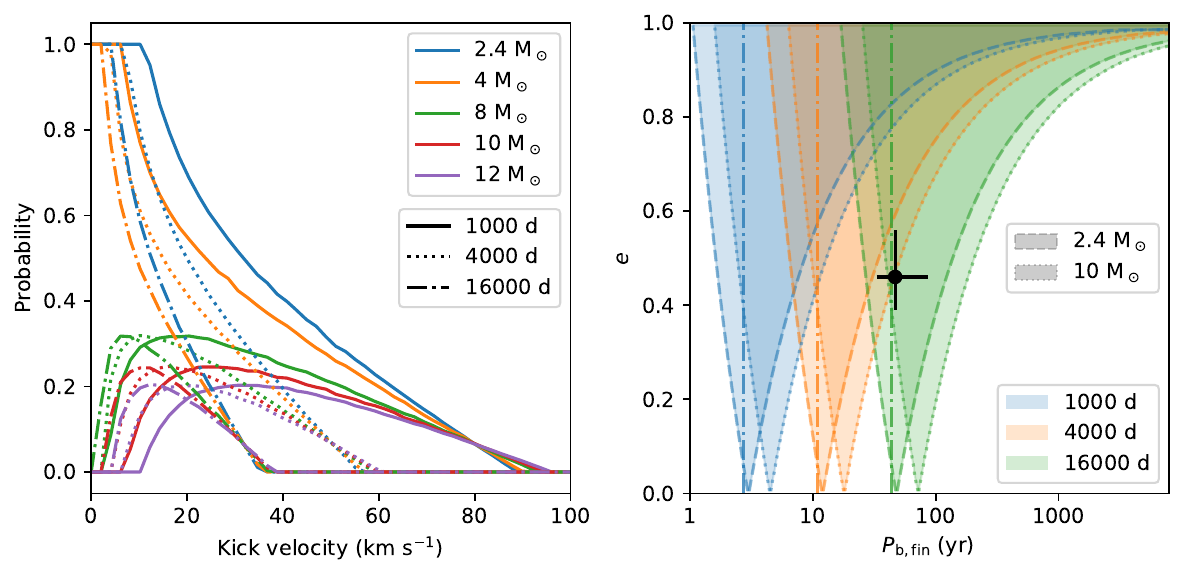}
    \caption{Results from supernova kick simulations. The left-hand panel shows the probability that the post-supernova binary remains bound as a function of kick velocity. These probabilities were computed with isotropic kicks for a range of pre-supernova primary masses for a Helium-star progenitor of 2.4 and 4\,M$_\odot$ and direct supernova explosions of 8, 10, and 12\,M$_\odot$ progenitors as well as pre-supernova orbital periods of 1000, 4000, and 16\,000\,d. The right-hand panel shows the resulting post-supernova orbital period $P_\mathrm{b,fin}$ and eccentricity for a range of pre-supernova masses and orbital periods. The dash-dotted vertical lines indicate the pre-supernova orbital periods with the wedges the resulting post-supernova parameters. The orbital constraints for PSR\,J0210+5845 are shown in black with error bars.}\label{fig:kicks}
\end{figure*}

We found that the orbit only remains bound for kicks with magnitudes below $v_\mathrm{kick}<96$\,km\,s$^{-1}$ and only if the kick direction is retrograde, i.e.\ opposite to the orbital velocity at the time of the supernova explosion. The kick velocity limit for which the system remains bound decreases as the initial orbit is wider, and has $v_\mathrm{kick}<37$\,km\,s$^{-1}$ for $P_\mathrm{b,ini}=16\,000$\,d, see Fig.\,\ref{fig:kicks}. The fraction of cases that remain bound (assuming isotropic kicks) is relatively low and decreases for higher progenitor masses.  Similarly, the probability for the system to remain bound improves for the alternative scenario where mass is lost from the system to form a Helium star. Repeating the calculations with a $M_\mathrm{p}=2.4$ to 4\,M$_\odot$ Helium star progenitor for the neutron star in a $P_\mathrm{b,ini}>1000$\,d pre-explosion orbit, the calculations show that the binary system remains bound for retrograde kicks with velocities below $v_\mathrm{kick}<90$\,km\,s$^{-1}$, but the probability increases to 100\% for velocities below $v_\mathrm{kick}<11$\,km\,s$^{-1}$, where also prograde orbits are possible.

If we consider a scenario with a less massive companion, the likelihood of the binary system surviving the supernova diminishes even further. Lower initial natal kicks to the neutron star are necessary to prevent disruption of the binary system.

The post-supernova orbits have a large range in possible orbital periods and eccentricities. Orbits that are initially wide are able to reproduce the observed orbital parameters determined with the spin frequency derivatives ($P_\mathrm{b}=47^{+40}_{-14}$\,yr, $e=0.46^{+0.10}_{-0.07}$). The post-supernova orbital period does not strongly depend on the progenitor mass or the eccentricity, as shown in Fig.\,\ref{fig:kicks}. We found that for the 8 to 12\,M$_\odot$ progenitor, the pre-supernova orbit would require orbital periods in the range of $P_\mathrm{b,ini}=2000$ to 40\,000\,d, while the Helium-star scenario with $M_\mathrm{p}=2.4$ to 4.0\,M$_\odot$ has compatible post-supernova orbits for $P_\mathrm{b,ini}=3000$ to 56\,000\,d.

As a result of the low natal kick velocities, the velocity imparted on the post-supernova binary centre-of-mass is also low. The formalism by \citet{Brandt95} predicts system velocities of $v_\mathrm{sys}=25$ to 45\,km\,s$^{-1}$ for a 10\,M$_\odot$ neutron star progenitor in a 1500\,d pre-supernova orbit for the range of kick velocities in which the binary remains bound. These velocities decrease for lower mass neutron star progenitors and wider pre-supernova orbits.

\section{Discussion and conclusions}\label{sec:discussion}
The LOFAR timing observations of PSR\,J0210+5845 reveal large deviations from typical spin-down behaviour of isolated pulsars that can be modelled by a spin frequency and a spin frequency derivative. We argue that these deviations are caused by binary motion of the pulsar in a wide, as yet unresolved, orbit around the B6\,V star 2MASS\,J02105640+5845176. This identification of 2MASS\,J02105640+5845176 as the binary companion of PSR\,J0210+5845 is based on the coincidence in celestial position, as well as the distance between the star and the pulsar, and the low probability for this coincidence to be due to random chance. 

The properties of PSR\,J0210+5845, its B6\,V binary companion, the orbital properties of this system and its location near the Galactic plane ($l=133\fdg10$, $b=-2\fdg54$) are consistent to those of high-mass binary pulsars. With this identification, PSR\,J0210+5845 becomes the seventh system with this classification, i.e.\ non-recycled pulsars ($P>0.01$\,s) with main-sequence star binary companions with masses in excess of $M_\mathrm{c}>1$\,M$_\odot$ \citep{Manchester05}. Compared to the other systems, PSR\,J0210+5845 stands out primarily due to the low mass of its binary companion. For those systems where companion masses have been reliably measured, the lowest mass is around $\sim8$\,M$_\odot$ (Table\,\ref{tab:hmbps}), while for the companion of PSR\,J0210+5845, we determined a mass of $3.5$ to $3.8$\,M$_\odot$. Similarly, among the high-mass X-ray binaries with measured masses, the lowest masses are around $\sim8$\,M$_\odot$ \citep[from the HMXB catalogue by][]{Fortin23}. The spin frequency derivatives determined from the timing of PSR\,J0210+5845 constrain the orbit to be wide ($P_\mathrm{b}=34$ to 88\,yr) with a moderate eccentricity of $e=0.39$ to 0.56. Of the high-mass binary pulsars, only PSR\,J2032+4127 is in a wide orbit of 46\,yr, though with a higher eccentricity of $e=0.96$ \citep{Lyne15}. All other systems have orbital periods below 2000\,d.

Due to the low mass of the companion, we found that for the binary to remain bound after the supernova explosion that formed PSR\,J0210+5845, a low velocity, retrograde natal kick is required ($v_\mathrm{kick}<96$\,km\,s$^{-1}$). This is true for both the direct collapse of a $>8$\,M$_\odot$ neutron star progenitor as well as the collapse of a Helium star. 

Low natal kicks of a few tens of km\,s$^{-1}$ are commonly attributed to electron-capture supernovae, where the rapid explosion does not allow for asymmetries to develop \citep[e.g.][]{Podsiadlowski04,Gessner18}. The progenitor mass range in which electron-capture supernovae occur is uncertain but estimated between 8 and 10\,M$_\odot$. Recently, \citet{Stevenson22} have postulated that electron-capture supernovae would create non-recycled pulsars in wide ($P_\mathrm{b}\ga10^4$\,d) and moderately eccentric ($e\sim0.7$) orbits. Their population synthesis predicts a distribution of post-supernova orbital periods and eccentricities matching the observed properties of PSR\,J0210+5845. Therefore,  PSR\,J0210+5845 is a possible candidate for this formation scenario.

Although in traditional Fe core-collapse supernovae, large natal kicks of hundreds of km\,s$^{-1}$ are predicted to be more common \citep[][and references therein]{Janka17}, observations show that lower natal kicks below 60\,km\,s$^{-1}$ are still possible \citep{Verbunt17}. Due to the broader range of progenitor masses leading to a core-collapse supernova that produces a neutron star \citep{Smartt09}, this scenario is not to be disregarded. 

The wide post-supernova orbit places constraints on the pre-supernova orbit, which we found requires to also be wide, with orbital periods ranging from 2000 to 56\,000\,d. This would argue against the formation scenario in which an episode of mass transfer between the neutron star progenitor and the binary companion removes the envelope of the progenitor, allowing it to explode as a Helium star. First, the current low mass of the companion does not allow the neutron star progenitor to have accreted much matter, and second, the wide pre-supernova orbit is not expected in this scenario where mass-transfer is required. However, through calculations of angular momentum pre- and post-interaction  and assuming no accretion by the companion, we found that under the ideal assumption of isotropic mass loss (e.g.\ see \citealt{Pols94}), orbital periods exceeding 4000 days are achievable for a Helium star. 

The orbital parameters indicate that the upcoming periastron passage is between 2030 and 2034. Continued timing observations around that time will significantly improve the orbital constraints. During periastron passage, the distance between the binary components is predicted to be $11.7^{+5.0}_{-1.5}$\,AU, which, using the approximation from \citet{Eggleton83} and assuming a mass ratio of 1.4/3.6, gives a Roche lobe radius of $R_\mathrm{L} = 756_{-101}^{+325}$\,R$_\odot$ \citep{Eggleton83}. Comparatively, a 3.6\,M$_\odot$ star has an approximate radius of 2.6\,R$_\odot$ \citep{Demircan91}, which indicates that there will be no Roche-lobe overflow.

At the nominal distance of 2.5\,kpc, the observed proper motion of 2MASS\,J02105640+5845176 corresponds to a transverse velocity of 14.5\,km\,s$^{-1}$. This velocity is the sum of the projected orbital velocity of the star around the binary centre-of-mass, the projected post-supernova system velocity of the binary centre-of-mass imparted on the system due to the supernova kick, and the projected component of any pre-supernova system velocity of the binary system. From the orbital constraints, we obtained a projected orbital velocity of the pulsar companion around the binary centre-of-mass of $3.1^{+0.5}_{-0.6}$\,km\,s$^{-1}$. Assuming a random direction of the post-supernova system velocity, the projected component will be less than 22\,km\,s$^{-1}$, indicating that the majority of the observed proper motion is due to the system velocity of the binary.

It remains to be seen if the orbital motion of the B6\,V star around the centre-of-mass of the binary system can be detected by {\it Gaia}. The orbital constraints based on the observed spin frequency derivatives of PSR\,J0210+5845 indicate that the projected acceleration due to orbital motion is small, of order 0.02\,mas\,yr$^{-2}$ at a distance of 2.5\,kpc, and the {\it Gaia} DR3 astrometric solution for position, proper motion and parallax reports no astrometric excess noise \citep{Gaia23}.

It has been proposed that numerous radio pulsars that are considered to be isolated, might in fact belong to exceptionally wide binary systems. \citet{Jones23} estimate that approximately 30\% of seemingly isolated pulsars with a measured $\ddot{f}$ could hide a binary with orbital period $<1000$ years. In this paper, we have demonstrated that even in these highly separated systems, the orbital motion of the pulsar can be measured through higher-order frequency derivatives. 

\begin{acknowledgements}
This paper is based (in part) on data obtained with the International LOFAR Telescope (ILT) under project codes LC9\_023, LC9\_041, LT10\_015 and LT14\_005. LOFAR \citep{vanHaarlem13} is the Low Frequency Array designed and constructed by ASTRON. It has observing, data processing, and data storage facilities in several countries, that are owned by various parties (each with their own funding sources), and that are collectively operated by the ILT foundation under a joint scientific policy. The ILT resources have benefitted from the following recent major funding sources: CNRS-INSU, Observatoire de Paris and Universit\'e d'Orl\'eans, France; BMBF, MIWF-NRW, MPG, Germany; Science Foundation Ireland (SFI), Department of Business, Enterprise and Innovation (DBEI), Ireland; NWO, The Netherlands; The Science and Technology Facilities Council, UK; Ministry of Science and Higher Education, Poland. This research was made possible by support from the Dutch National Science Agenda, NWA Startimpuls – 400.17.608. IVY and JC acknowledge support by the Spanish Ministry of Science under grant PID2020-120323GB-I00.The INT is operated on the island of La Palma by the Isaac Newton Group of Telescopes in the Spanish Observatorio del Roque de los Muchachos of the Instituto de Astrof\'{i}sica de Canarias.

\end{acknowledgements}

\bibliographystyle{aa}
\bibliography{LOTAAS}

\end{document}